# The antibunching effect of polaritons in a microcavity dimer

T.A.Khudaiberganov[1]

[1]Department of Physics and Applied Mathematics,
Vladimir State University named after A. G. and N. G. Stoletovs, 87 Gorkogo, Vladimir, 600000, Russia
E-mail thomasheisenberg@mail.ru

**Summary:** We are consistency considered two cases. Firstly, we consider exciton-photon statistic radiation from pillar microcavity. We obtained a photon antibunching and small polariton antibunching. Secondly, we use two strong-coupled pillar microcavities to achieve pronounced polariton antibunching. We observed the polariton blockade effect when use a polarion dimer.

## 1. Introduction

Studying of the quantum statistical properties light-matter nonlinear systems is a great interesting in quantum optics. The harvest of which is a highly desirable for quantum information, for example in the make of a single-photons sources. Usually, interesting quantum properties (antibunching, photon blockade) of light are achieved highly nonlinearity comparable to the relaxation parameters in the system. Unconventional quantum statistics implies weakly nonlinear systems (U<<γ), which is a more reality a value of a nonlinearity in the condensed matter physics [1,2].

We considered the exciton-polariton system formed in coupled 0D pillar microcavities – polariton dimer [3] with typical for this system weakly nonlinearity U≈0.01γ, see fig.1. Quantum statistic light from microcavity has already been studied [4,5], the antibunching, bunching and giant bunching phenomena have been discovered.

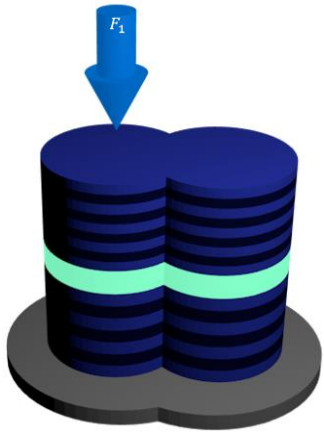

**Fig. 1.** Sketh a polariton dimer with a laser pump into the first micropillar.

## 2. Polariton dimer model

The following Hamiltonian describes system coupled anharmonic oscillators under laser pump energy (in the rotate wave approximation):

$$H = \Delta_1 \hat{a}_1^+ \hat{a}_1 + \Delta_2 \hat{a}_2^+ \hat{a}_2 + g_{12}\hat{a}_1^+ \hat{a}_2 + g_{21}\hat{a}_2^+ \hat{a}_1 + \frac{U}{2}\hat{a}_1^+\hat{a}_1^+\hat{a}_1\hat{a}_1 + \frac{U}{2}\hat{a}_2^+\hat{a}_2^+\hat{a}_2\hat{a}_2 \quad (1a)$$

Here $\hat{a}(\hat{a}^+)$ annihilates (creates) bosons operators; $\Delta = \omega_i - \omega_L$ are the cavity detunings, $\omega_L$ - frequency driving fields $F_i$, which we will set equal to each other; $\omega_i$ amplitudes of the driving fields; $U_i$ are the Kerr nonlinearity parameters; $g_{ij} = g_{ji}$ are coupling constant between i- cavity and j- cavity (for polaritons is rabi energy, for coupled microcavity is hopping amplitude between the two cavities.)

The Hamiltonian (1a) can be model of the coupled microcavities and model an exciton-polaritons with the some differents

$$\hat{H}_C = \hbar\Delta_{ph}\hat{\phi}^+\hat{\phi} - \hbar\Delta_{ex}\hat{\chi}^+\hat{\chi} + \hbar\omega_R(\hat{\chi}^+\hat{\phi} + \hat{\phi}^+\hat{\chi}) + \hbar\alpha\hat{\chi}^{+2}\hat{\chi}^2 + \hbar E_d(\hat{\phi}^+ + \hat{\phi}). (1b)$$

Master equation by a matrix density for Hamiltonian (1),

$$\frac{d\rho}{dt} = -i[H_+^{eff},\rho] + \gamma_1 D[\hat{a}_1] + \gamma_2 D[\hat{a}_2] \quad (2)$$

Where we introduce the follow dissipators
$D[\hat{a}_1] = \hat{a}_1\rho\hat{a}_1^+ - \frac{1}{2}[\hat{a}_1^+\hat{a}_1,\rho]_+$, $\quad D[\hat{a}_2] = \hat{a}_2\rho\hat{a}_2^+ - \frac{1}{2}[\hat{a}_2^+\hat{a}_2,\rho]_+$.

## 3. Results

### 3.1. Exciton-polaritons

We can see pronounced antibunching of the photon field in the region $\Delta = -\Omega$, see fig.1 from [5]. While a small antibunching of polaritons is near the lower polariton branch resonance (shaded line), see fig.2.$\omega_L$

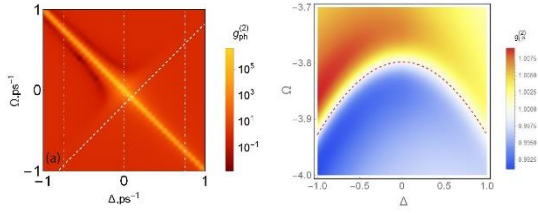

**Fig.2.** The second order correlation function of light from pillar micocavity for various values of the driving detuning and the exciton-photon detuning [5].

The statistics of exciton field is almost coherent, while the photon field statistics reaches s strong antibunching. The quantum statistics of lower branch polaritons (LP-polaritons) combines the statistics of photons and excitons, and due to the predominance of the exciton component, the quantum statistics of LP-polaritons is close to the statistics of excitons i.e. to coherent statistics. The statistics of polaritons has been study earlier in [6], as well as in the recent work [7]. To achieve unconventional statistics (for small values of the nonlinearity U), we have to set the stationary photon (or polaritons) field around zero i.e. almost vacuum state (average of number bosons approximately equal to $10^{-6}$) or greatly increase the pumping [5]. In a pillar microcavity non-classical statistics arises not due to quantum interference [7], in opposite to the usual convention non-classical statistics, but due to the fact that one oscillator slaved [5] the quantum statistics of another oscillator. Namely, the small antibunching of the exciton field leads to antibunching of the photon field.

### 3.2. A polariton dimer

The second order correlation function we can define as,

$$g_1^{(2)} = \frac{\sum_{n,m} n(n-1)\rho_{n,n,m,m}}{(\sum_{n,m} n\rho_{n,n,m,m})^2} \approx \frac{\rho_{2,2,0,0}}{\rho_{1,1,0,0}^2} \quad (3)$$

Here $\rho_{n,n',m,m'} = \langle n'm'|\rho|nm\rangle$

On the Fig. 3 shows the second-order correlation function of the first microresonator at for two values of the nonlinearity parameter for (a-b) U = 0.001γ and (c-d) U = 0.01γ in the detuning range $\Delta_1$ and $\Delta_2$. The minimum value for the system under consideration is $g_{min}^{(2)} \sim 10^{-3}$. The system of microresonators is asymmetric. All the pump energy to the first microresonator goes to the second microresonator. In this case, the population of the number of photons in the second microresonator exceeds the population of the number of photons in the first one. In this case, the radiation statistics of the second microcavity remains coherent, and the statistics of the radiation of the first microcavity shows nonclassicality in the region of positive detuning of the second microcavity.

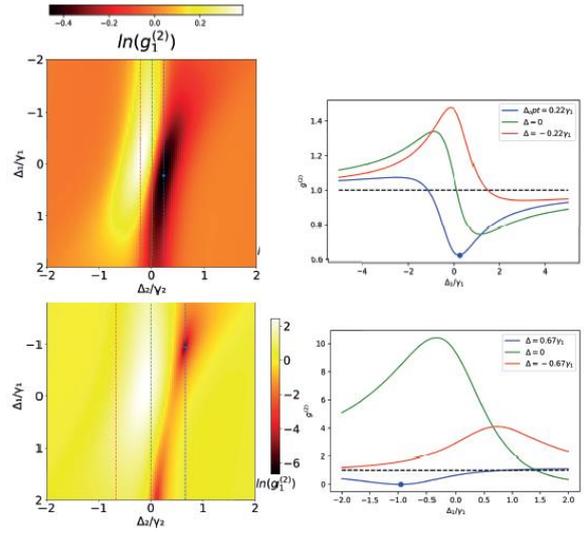

**Fig.3.** The second order correlation function of the lower-polaritons in dependency detunings for (a-b) U = 0.001γ and (c-d) U = 0.01γ.

The mechanism of non-traditional quantum blockade in a two-mode system is based on non-destructive quantum interference of quantum trajectories of population of states at a small value of the nonlinearity parameter U< γ and low pump power $F_p \sim γ$. The mechanism proposed for an asymmetric two-mode exciton-polariton system and operates at low values of the nonlinearity parameter and high pump power.

On the Fig.4 are shows the behavior function $g_1^{(2)}$ near the region where the effect of polariton blockade and matrix elements $\rho_{2,0,2,0}$ and $\rho_{1,0,1,0}$. the matrix element $\rho_{2,0,2,0}$ experiences a minimum at the minimum point of the function $g_1^{(2)}$.

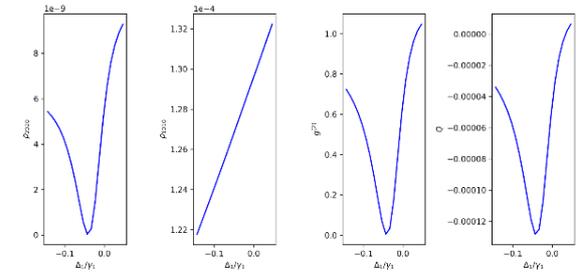

**Fig.4.** Matrix density elements (c) second order and (d) Q-function.

The coupled microcavities model [2,8] in frame lower polarioton (LP) truncated basis approximation [9] with parameters [10]: g = 5.62γ, $U_1 = U_2 = U_{LP} = C_{LP}^4 U_{ex} = 0.01\gamma$. And pump is only on the first mode, $F_1 = \gamma_1, F_2 = 0$

We used a numerical method for solving the master equation (2) for the density matrix for the Hamiltonian (1). For this, we used the truncated Liuvillian-Fock bases space where only states $|nm\rangle$ with n+m ≤ $N_{max}$. In our case we take $N_{max}$=10. We used the qutip library by python.

## 4. Conclusions

We suggest instead of using a single microresonator used two the strong-coupled microcavities for obtained non-classical states of a polaritons. We approach use unconventiona polariton blockade. We numeric obtained resonace condititon for the polaritons quantum blockade.

## Acknowledgements

The research was carried out within the state assignment in the field of scientific activity of the Ministry of Science and Higher Education of the Russian Federation (theme FZUN-2020-0013, state assignment of VlSU). The study was carried out using the equipment of the interregional multispecialty and interdisciplinary center for the collective usage of promising and competitive technologies in the areas of development and application in industry/mechanical engineering of domestic achievements in the field of nanotechnology (Agreement No. 075-15-2021-692 of August 5, 2021).